# Place & Play SERS: sample collection and preparation-free surface-enhanced Raman spectroscopy

Yasutaka Kitahama,[a,b] Pablo Martinez Pancorbo,[a] Hiroki Segawa,[c] Machiko Marumi,[a] Ting-Hui Xiao,[a,b,d] Kotaro Hiramatsu,[a] William Yang,[e] and Keisuke Goda*[a,b,d,f,g]

a  Department of Chemistry, The University of Tokyo, Tokyo 113-0033, Japan. E-mail: goda@chem.s.u-tokyo.ac.jp
b  LucasLand, Co. Ltd, Tokyo 101-0052, Japan
c  Third Department of Forensic Science, National Research Institute of Police Science, Chiba 277-0882, Japan
d  Institute for Quantum Life Science, National Institute for Quantum and Radiological Science and Technology, Chiba 263-8555, Japan
e  BaySpec, San Jose, CA 95131, USA
f  Institute of Technological Sciences, Wuhan University, Hubei 430072, China
g  Department of Bioengineering, University of California, Los Angeles, California 90095, USA

**Abstract**
The ability to perform sensitive, real-time, in situ, multiplex chemical analysis is indispensable for diverse applications such as human health monitoring, food safety testing, forensic analysis, environmental sensing, and homeland security. Surface-enhanced Raman spectroscopy (SERS) is an effective tool to offer the ability by virtue of its high sensitivity and rapid label-free signal detection as well as the availability of portable Raman spectrometers. Unfortunately, the practical utility of SERS is limited because it generally requires sample collection and preparation, namely, collecting a sample from an object of interest and placing the sample on top of a SERS substrate to perform a SERS measurement. In fact, not all analytes can satisfy this requirement because the sample collection and preparation process may be undesirable, laborious, difficult, dangerous, costly, or time-consuming. Here we introduce "Place & Play SERS" based on an ultrathin, flexible, stretchable, adhesive, biointegratable gold-deposited polyvinyl alcohol (PVA) nanomesh substrate that enables placing the substrate on top of an object of interest and performing a SERS measurement of the object by epi-excitation without the need for touching, destroying, and sampling it. Specifically, we characterized the sensitivity of the gold/PVA nanomesh substrate in the Place & Play SERS measurement scheme and then used the scheme to conduct SERS measurements of both wet and dry objects under nearly real-world conditions. To show the practical utility of Place & Play SERS, we demonstrated two examples of its application: food safety testing and forensic analysis. Our results firmly verified the new measurement scheme of SERS and are expected to extend the potential of SERS by opening up untapped applications of sensitive, real-time, in situ multiplex chemical analysis.

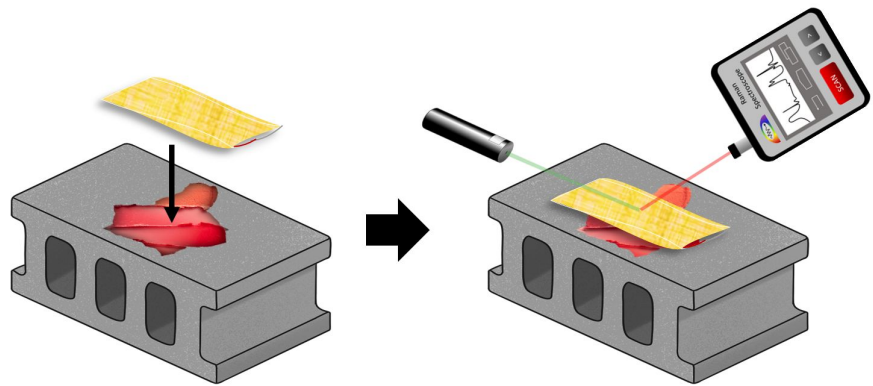

## 1. INTRODUCTION

The ability to perform sensitive, real-time, in situ, multiplex chemical analysis is indispensable for diverse applications such as human health monitoring, food safety testing, forensic analysis, environmental sensing, and homeland security.[1–5] Surface-enhanced Raman spectroscopy (SERS), a method for significantly enhancing inherently weak Raman scattering by localized surface plasmon resonance and measuring the enhanced Raman scattering signal for nondestructive structural analysis of samples,[6–25] is an effective tool to provide the ability by virtue of its high sensitivity (even down to single molecules) and rapid label-free signal detection (down to ~1 s or less) as well as the availability of portable Raman spectrometers (down to the passport size and 0.5 kg).[24–26] SERS is advantageous over other analytical methods such as mass spectrometry, gas chromatography, liquid chromatography, laser-indued breakdown spectroscopy, and fluorescence detection because these methods require bulky and costly equipment, are time-consuming in sample collection and preparation, or lack the capability of nondestructive, multiplex detection.[1–4] In fact, SERS has been widely used in trace detection of hazardous chemicals, pesticide residues, microplastics, glucose in whole blood, viruses of infectious diseases, and controlled drugs.[7,24,25]

Unfortunately, the practical utility of SERS is limited because it generally requires sample collection and preparation, namely, collecting a sample from an object of interest and placing the sample on top of a SERS substrate to perform a SERS measurement (Fig. 1a).[8] In fact, not all analytes can satisfy this requirement because the sample collection and preparation process may be undesirable, laborious, difficult, dangerous, costly, or time-consuming.[1–5,27] For example, a dry sample must be collected and dissolved into a liquid sample, which is mixed with a colloidal suspension or dropped onto a SERS substrate for a SERS measurement.[8] Also, the sample collection and preparation process itself could alter the chemical structure of the sample, making it unable to perform accurate in situ chemical analysis.[4,5,27] Moreover, it is difficult to collect potentially hazardous materials and conduct sample preparation before identifying what they are.[1–5,27] Finally, hard materials such as solid rocks and semiconductor materials are difficult to sample.[4,5,26,27] It would be highly beneficial to SERS users and further expand the application range of SERS if the sample collection and preparation process could be skipped or alleviated.

In this paper, we introduce "Place & Play SERS" based on an ultrathin, flexible, stretchable, adhesive, biointegratable gold-deposited polyvinyl alcohol (PVA) nanomesh substrate previously demonstrated by our group[13], a method that enables placing the substrate on top of an object of interest and performing a SERS measurement of the object by epi-excitation without the need for touching, destroying, and sampling it (Fig. 1b). The gold/PVA nanomesh substrate (2 – 10 μm) is thinner than previous SERS substrates based on a similar concept (93 – 280 μm)[14–16] and can be attached to virtually any target surface via dissolution of PVA as an adhesive. The gold/PVA nanomesh substrate shows a high SERS signal enhancement of ~$10^8$, high sensitivity down to ~10 nM of rhodamine 6G (R6G), and high SERS measurement reproducibility (i.e., spectrum-to-spectrum consistency) due to a high density of plasmonic hot spots in three dimensions that originate from the nanometer-sized gaps of gold nanofibres in the gold/PVA nanomesh as well as the edges of the gold nanofibres.[13] Specifically, we characterized the sensitivity of the gold/PVA nanomesh substrate in the Place & Play SERS measurement scheme and then used the scheme to conduct SERS measurements of both wet and dry objects under nearly real-world conditions. For SERS measurements of dry objects, we only needed to spray them with water before placing the SERS substrate on the surface of the objects. To show the practical utility of Place & Play SERS, we demonstrated two practical applications: (1) food safety testing by detecting a biocide (5-chloro-2-methyl-4-isothiazolin-3-one; CMIT) on an orange peel, (2) forensic analysis by detecting dry controlled drugs (etizolam, 3,4-methylenedioxymethamphetamine; MDMA) on an aluminium foil. Our results firmly verified the new measurement scheme of SERS and are expected to extend the potential of SERS by opening up untapped applications of sensitive, real-time, in situ multiplex chemical analysis.

## 2. MATERIALS AND METHODS

### 2.1 Materials
We used PVA, CMIT (14% aqueous solution), and etizolam purchased from Wako Chemicals. We used R6G purchased from Sigma Aldrich. We used MDMA stored as a methanol solution in our laboratory. Distilled water was collected from Milli-Q (Merck Millipore) with vent filters MPK01 and Q-POD1 at 15.0 MΩ and 14.6 °C.

### 2.2 Fabrication
The fabrication process of the gold/PVA nanomesh substrate was described in our previous report with modifications.[13] In brief, a mesh of PVA nanofibres or PVA nanomesh was fabricated by electrospinning (MECC, NEX-101) at 20 kV for 1 h with 8 wt% PVA aqueous solution flowing at a rate of 1 mL/h using a microfluidic pump. The thickness of the PVA nanomesh was measured with a thickness gauge (Mitutoyo, 547-401). Then, a 150-nm-thick gold layer was thermally deposited on the PVA nanomesh at a flow rate of 0.15 nm/sec under a high vacuum (less than $5 \times 10^{-4}$ Pa) to produce gold/PVA nanomesh. We used a scanning electron microscope (SEM) (Hitachi, Regulus 8230) to evaluate the fabricated gold/PVA nanomesh.

### 2.3 Physical characterization of the gold/PVA nanomesh substrate
To characterize the gold/PVA nanomesh substrate, we imaged it using the SEM. Fig. 2 shows an image of the gold/PVA nanomesh substrate (Fig. 2a) and SEM images of the deposited and undeposited sides of the gold/PVA nanomesh substrate (Fig. 2b, Fig. 2c). On the deposited side (top surface), gold nanofibres with diameters of several hundred nanometers are evident, revealing that the surface of the gold nanofibres is rough (Fig. 2b). On the other hand, as shown in Fig. 2c, the gold layer was not deposited on all of the PVA nanofibres. The PVA nanofibres on the undeposited side (bottom surface) of the gold/PVA nanomesh substrate show that the surface of the PVA nanofibres is smooth and the gold was deposited on the edges of the PVA nanofibres (Fig. 2c). Fig. 2d shows the gold nanofibres on the undeposited side (bottom surface) of the gold/PVA nanomesh substrate after removing PVA by placing the substrate onto the surface of an object of interest, optionally spraying it with water if the target analyte is not an aqueous solution, and dissolving PVA into water to form gutter-like structures.

### 2.4 SERS measurements
Three different procedures were used to obtain SERS spectra of analytes with the gold/PVA nanomesh substrate. (1) The gold/PVA nanomesh substrate was attached to the target surface, which was sprayed with water to obtain pure gold nanofibres via dissolving PVA into water. An analyte solution (3 μL) was dropped on the deposited side (top surface) of the substrate on the target surface (Fig. 1a, conventional SERS). (2) The gold/PVA nanomesh substrate was placed on an analyte solution (3 μL) on top of the target surface (Fig. 1b, Place & Play SERS). The analyte was adsorbed on the undeposited side (bottom surface) of the gold/PVA nanomesh substrate. (3) An analyte solution (3 μL) was dried on the target surface and sprayed with water. The wetted analyte was covered with the gold/PVA nanomesh substrate (Fig. 1b, Place & Play SERS) and adsorbed on the undeposited side (bottom surface) of the gold/PVA nanomesh substrate.

### 2.5 Spectrometers
SERS spectra were acquired using a wearable Raman spectrometer (WSERS-785, BaySpec) with a spectral resolution of 16–18 cm$^{-1}$ by excitation at 785 nm and 20 mW (tunable from 10 to 80 mW using a computer) for an integration time of 2.5 s (tunable from 1 ms to 120 s). The obtained spectra were automatically corrected by baseline subtraction. The SERS spectra of the controlled drugs were obtained using a laser Raman microscope (RAMANforce, Nanophoton Corp.) by excitation at 785 nm and 10 mW through a 20× objective lens (NA 0.45) for an exposure time of 1 s with 16 accumulations. The obtained spectra were manually corrected by subtracting the background (Raman spectrum from the gold/PVA nanomesh substrate itself without samples). To investigate SERS sensitivity on the gold/PVA nanomesh substrate with various substrate thickness values, a SERS signal intensity at a particular Raman shift value was obtained using an RM2000 confocal Raman microscope (InVia, Renishaw PLC, England) by excitation at 785 nm and 10 mW through a 50× objective lens with a long working distance (NA 0.55) for an exposure time of 1 s with 20 accumulations. In all SERS measurements regardless of whether they were conducted on the top or bottom surface of the gold/PVA nanomesh substrate, the distance between the substrate and the wearable Raman spectrometer was set to be 4.0 mm.

## 3. RESULTS AND DISCUSSION

### 3.1 Demonstration of Place & Play SERS on various surfaces

To demonstrate Place & Play SERS on virtually any surface, we attached the gold/PVA nanomesh substrate to various surfaces and measured SERS spectra of R6G (1 mM) on the surfaces using the wearable Raman spectrometer (Fig. 3a – Fig. 3f). In the demonstrations below, the gold/PVA nanomesh substrate was optionally sprayed with water to dissolve PVA from the substrate so that the substrate could be attached to the target surface. If the object of interest was a solution (e.g., water, ethanol), this step was not required. First, to show Place & Play SERS on water-repellent surfaces, we conducted SERS measurements of R6G on a polystyrene dish (Fig. 3a) and a nitrile butadiene rubber (NBR) glove (Fig. 3b). It is evident from the figures that the SERS spectrum of R6G from the bottom surface of the substrate in the Place & Play SERS measurement scheme shown in Fig. 1b is nearly identical in spectral shape and intensity to that from the top surface of the substrate in the conventional SERS measurement scheme shown in Fig. 1a, validating Place & Play SERS. Furthermore, the figures also show that a trace of R6G after drying was detectable by spraying the dried R6G with water and placing the substrate on top of it. Second, to show Place & Play SERS on water-absorbent surfaces, we performed SERS measurements of R6G on a polyester cloth (Fig. 3c), a concrete brick (Fig. 3d), a face mask (Fig. 3e), and a paper towel (Fig. 3f). Like the results shown in Fig. 3a and Fig. 3b, it is evident from the figures that the SERS spectrum of R6G from the bottom surface of the substrate in the Place & Play SERS measurement scheme shown in Fig. 1b is nearly identical in spectral shape and intensity to that from the top surface of the substrate in the conventional SERS measurement scheme shown in Fig. 1a, validating Place & Play SERS. Unlike the results shown in Fig. 3a and Fig. 3b, a trace of R6G after drying was difficult to detect despite spraying the dried R6G with water. This is because R6G was absorbed into the materials. It is noteworthy that the SERS spectrum of R6G was obtained by wetting the polyester cloth with water, presumably because R6G penetrated between the polyester fibres, not into the fibres, and did not flow out when it was sprayed with water. These results clearly demonstrated Place & Play SERS under practical conditions in real time in situ.

### 3.2 Sensitivity of Place & Play SERS

To compare the conventional SERS and Place & Play SERS measurement schemes in terms of sensitivity, we performed SERS measurements of R6G on a glass plate using the top and bottom surfaces of the gold/PVA nanomesh substrate under the same conditions. The limit of detection of R6G on the top surface of the gold/PVA nanomesh substrate was found to be 10 nM and 100 μM from its SERS spectra with an RM2000 confocal Raman microscope through a 50× objective lens (NA = 0.75)[13] and the wearable Raman spectrometer (Fig. S1 in the Supporting Information), respectively. Fig. 4a and Fig. 4b show the SERS peak intensity values of R6G aqueous solutions at various concentrations at 775 cm$^{-1}$, which is an isolated peak of R6G and hardly influenced by its neighbouring peaks and background, from the top and bottom surfaces of the gold/PVA nanomesh, respectively.

The figures show that while the sensitivities of the conventional SERS and Place & Play SERS measurement schemes were comparable at high concentrations of R6G, they differed at low concentrations. Since the SERS intensity values at low concentrations were nearly independent of the substrate thickness in the Place & Play SERS measurement scheme, optical excitation from the above to the top surface does not account for its lower sensitivity at low concentrations compared with that of the conventional SERS measurement scheme. Instead, a large concentration of the dissolved PVA is presumably responsible for competing with R6G at low concentrations, resulting in a reduced sensitivity on the bottom surface while the sensitivity is retained on the top surface which is devoid of PVA.[28] These results indicate that the sensitivity of Place & Play SERS is comparable to that of conventional SERS except for very low concentrations.

### 3.3 Dependence of the sensitivity of Place & Play SERS on the substrate thickness

To evaluate the dependence of the sensitivity of Place & Play SERS on the thickness of the gold/PVA nanomesh substrate, we performed SERS measurements of R6G on the top and bottom surfaces of the substrate with various thickness values. Like the results in Fig. 4a and Fig. 4b, the SERS peak intensity values of R6G at 775 $cm^{-1}$ were used for the evaluation. As shown in Fig. 5a (top surface), thicker substrates tend to exhibit stronger SERS intensity values at high concentrations because the coverage of R6G molecules on hot spots increases with the substrate thickness. On the other hand, as shown in Fig. 5b (bottom surface), this trend is reversed because thinner substrates tend to allow the analyte to penetrate through the substrate more easily. It is worthwhile to note that error bars in Fig. 5a are larger than those in Fig. 5b. This may be due to an inhomogeneous spatial evaporation process, such as the coffee-ring effect,[29] on the top surface of the substrate.

### 3.4 Food safety testing with Place & Play SERS

To show the practical utility of Place & Play SERS, we demonstrated food safety testing with the gold/PVA nanomesh substrate. Specifically, we performed SERS measurements of a biocide, CMIT, on an orange peel. Here, biocides are chemical substances that act against yeast, bacteria, and fungi and have been applied to food additives as preservatives for plants and fruits in post-harvest.[30-32] CMIT is a class of heterocycles used as a biocide and is also used in many household products such as cosmetics and antiseptics. However, it causes chemical burns at high concentrations while "rinse-off" and "leave-on" cosmetic products are acceptable at low concentrations of 15 ppm and 7.5 ppm, respectively.[33] Fig. 6 shows the SERS spectra of CMIT aqueous solutions at various concentrations (0.14%, 1.4%, 14%) on an orange peel using the gold/PVA nanomesh and wearable Raman spectrometer. The SERS peaks that appeared at about 720 $cm^{-1}$, 940 $cm^{-1}$, and 1340 $cm^{-1}$ were tentatively assigned to N–$CH_3$ stretch, C–C skeletal stretch, and C–H deformation vibrational modes.[34,35] These vibrational modes were detectable at concentrations down to 1400 ppm (0.14%) even with the wearable Raman spectrometer. These results show the excellent performance of Place & Play SERS as opposed to the inability of spontaneous Raman spectroscopy to detect even the high concentration (14%) of CMIT on the orange peel as indicated by the agreement between the spontaneous Raman spectra of 14% CMIT on the orange peel and the orange peel itself (Fig. 6). Here, the peaks in these spontaneous Raman spectra are attributed to carotenoids in fruits,[36–38] with the peaks at 1525 $cm^{-1}$, 1150 $cm^{-1}$, and 1000 $cm^{-1}$ assignable to the in-phase $\upsilon$(C–C), $\upsilon$(C–C) stretching vibrations of the polyene chain[39] and the in-phase rocking modes of the $CH_3$ groups attached to the polyene chain,[40,41] respectively. These results firmly support the power of Place & Play SERS in nondestructive, real-time, in situ Raman-based food safety testing.

### 3.5 Forensic analysis with Place & Play SERS

To further show the practical utility of Place & Play SERS, we demonstrated forensic analysis with the gold/PVA nanomesh substrate. Specifically, we conducted SERS measurements of two types of controlled drugs, etizolam (a type of benzodiazepine used for short-term treatment of insomnia and anxiety disorders) and MDMA (a synthetic stimulant and psychedelic also known as "ecstasy" or "molly"), on an aluminium foil in dry conditions using the gold/PVA nanomesh substrate and Raman microscope. In this application, the Place & Play SERS measurement scheme is highly effective because it closely resembles the practical situation of controlled-drug detection in potential crime scenes such as when the target surface is frequently touched by a hand contaminated with the controlled drugs, compared with the previous SERS measurements that the sample was dropped on the gold/PVA nanomesh substrate.[13] As shown in Fig. 7a, the prominent peaks that appeared at about 450 cm$^{-1}$, 650 cm$^{-1}$, 1050 cm$^{-1}$, and 1500 cm$^{-1}$ agree with those previously reported.[42,43] These peaks were detectable at trace concentrations down to 0.1 mM after wetting with water (Fig. 7b) and down to 0.01 mM in an ethanol solution (Fig. 7b). Notably, the SERS spectrum of etizolam was obtained from the ethanol solution with the gold/PVA nanomesh containing PVA, which is insoluble in organic solvents except for dimethyl sulfoxide (DMSO). Likewise, as shown in Fig. 7c, the main peaks that appeared at 525 cm$^{-1}$, 715 cm$^{-1}$, 810 cm$^{-1}$, 1245 cm$^{-1}$, 1365 cm$^{-1}$, 1435 cm$^{-1}$, and 1490 cm$^{-1}$ agree with those of silver colloids excited at 514 nm, but not at 633 nm and 1064 nm.[44] These SERS peaks were detectable at trace concentrations down to 52 µM after wetting with water (Fig. 7c) and at 0.52 mM in a methanol solution (Fig. 7d), while no Raman peak of MDMA at a trace concentration of 0.52 mM on a silicon substrate was obtained with the same laser Raman microscope.[13] These results indicate the superior capability of Place & Play SERS in nondestructive, real-time, in situ Raman-based forensic analysis.

**CONCLUSIONS**

In this paper, we have demonstrated sample collection and preparation-free SERS, namely "Place & Play SERS". This new type of SERS has been made possible by employing an ultrathin, flexible, stretchable, adhesive, biointegratable gold-deposited polyvinyl alcohol (PVA) nanomesh substrate. It enables placing the substrate on top of an object of interest and performing a SERS measurement of the object by epi-excitation without the need for touching, destroying, and sampling it. Specifically, we have used the substrate to perform SERS measurements of both wet and dry objects under nearly real-world conditions. Furthermore, to show the practical utility of Place & Play SERS, we have used the substrate to demonstrate food safety testing by detecting a biocide, CMIT, and forensic analysis by detecting dry controlled drugs, etizolam and MDMA, in a nondestructive, real-time, and in situ measurement manner. Our results indicate that Place & Play SERS holds great promise for diverse biomedical applications.

**AUTHOR CONTRIBUTIONS**

K.G. conceived the concept of the work. Y.K. mainly conducted the experiments, obtained the data, and analyzed the results. P.M.P. and M.M. helped Y.K. with the fabrication of the gold/PVA nanomesh substrate. H.S. performed the forensic analysis experiments. W.Y. provided the wearable Raman spectrometer and optimized its performance. T.H.X., K.H., and K.G. supervised the work. K.G. acquired funding. All authors have approved the final version of the manuscript.

**CONFLICT OF INTEREST**

K.G. is a shareholder of LucasLand. W.Y. is a shareholder of BaySpec. T.-H.X., K.H., and K.G. hold a pending patent for the gold/PVA nanomesh SERS substrate.

**ACKNOWLEDGEMENTS**

This research was mainly supported by MEXT Quantum Leap Flagship Program (JPMXS0120330644), Mitsubishi UFJ Technology Development Foundation, and UTokyo IPC and partly supported by JSPS Core-to-Core Program and MEXT Advanced Research Infrastructure for Materials and Nanotechnology in Japan (ARIM) (JPMXP1222UT1039).

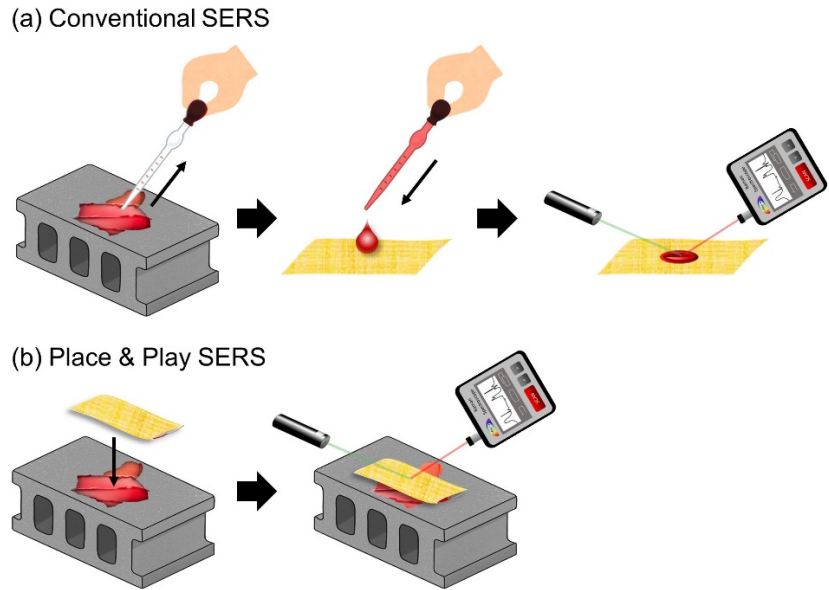

**Fig. 1** Differences in measurement scheme between conventional SERS and Place & Play SERS. (a) Conventional SERS measurement scheme. (b) Place & Play SERS measurement scheme.

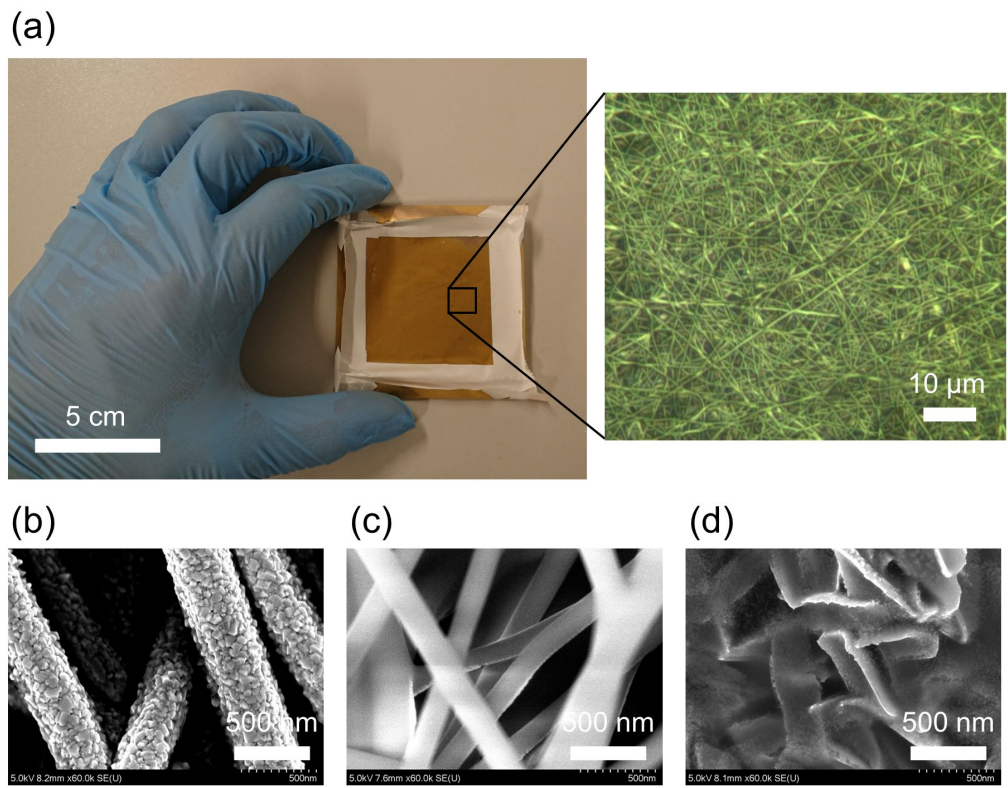

**Fig. 2** Gold/PVA nanomesh SERS substrate. (a) Picture of the substrate. The inset shows a 100× optical microscopy image of the gold/PVA nanomesh substrate. (b) SEM image of the deposited side of the substrate. (c) SEM image of the undeposited side of the substrate containing PVA. (d) SEM image of the undeposited side of the substrate after removing PVA.

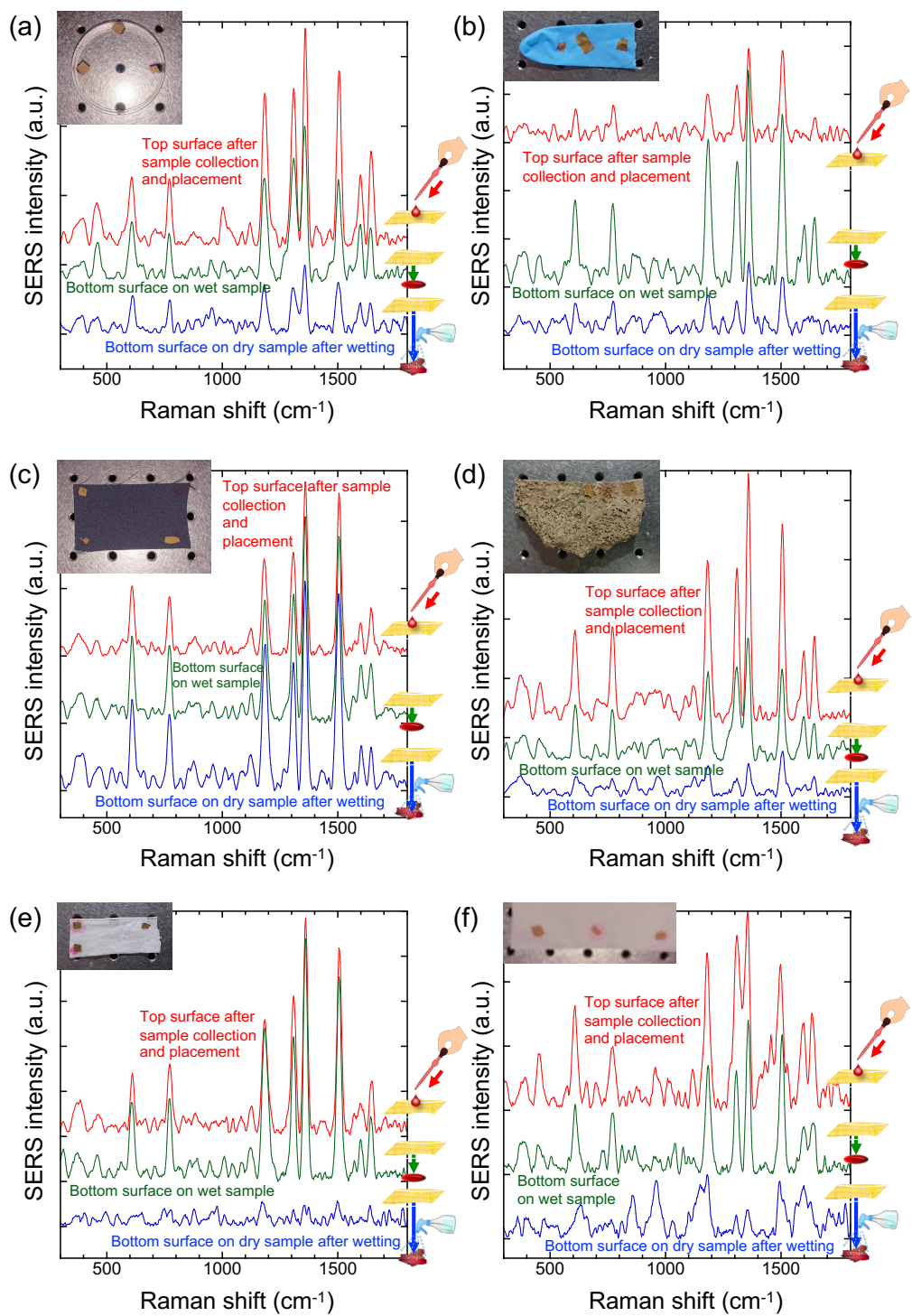

**Fig. 3** SERS spectra of R6G (1 mM) on the top and bottom surfaces of the gold/PVA nanomesh substrate placed on various surfaces: (a) polystyrene dish, (b) NBR glove, (c) polyester cloth, (d) concrete brick, (e) face mask, (f) paper towel. Insets show pictures of the substrate and various materials. All the analytes were optically interrogated from the top surface of the substrate at an excitation power of 20 mW at 785 nm for a duration of 2.5 s. Note that the SERS spectra of R6G on the top and bottom surfaces of the substrate were separately obtained.

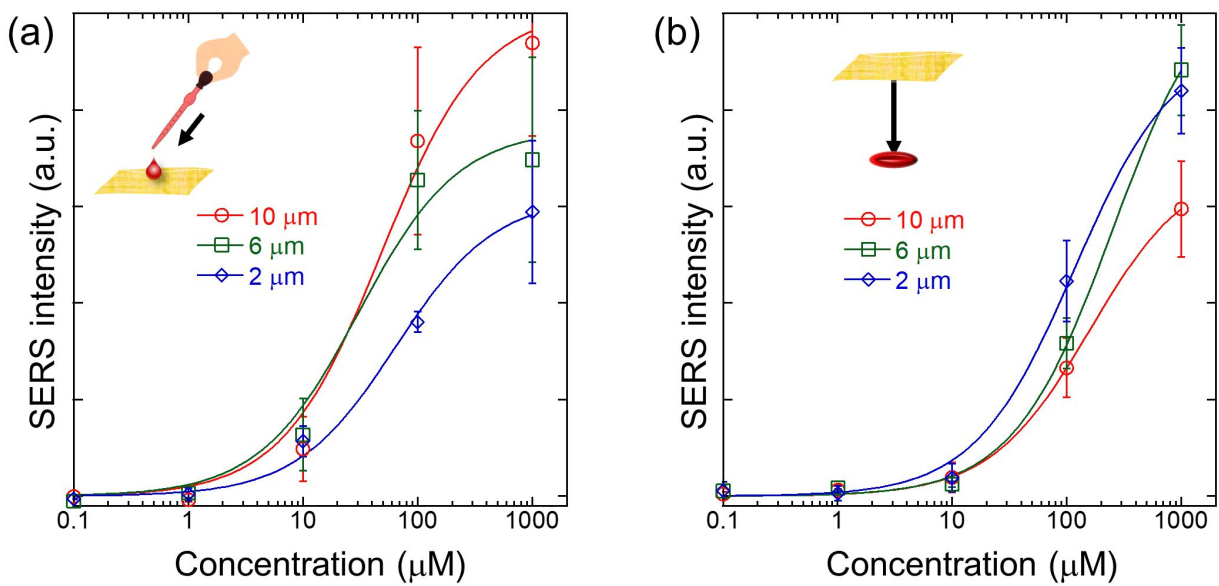

**Fig. 4** SERS peak intensity of an R6G aqueous solution at 775 cm$^{-1}$ obtained on the gold/PVA nanomesh substrates with thickness values of 10 μm, 6 μm, and 2 μm at various concentrations. Error bars indicate the standard deviations of repeated measurements (n = 3). Polynomial fitting was applied to visualize trends. (a) On the top surface of the substrate. (b) On the bottom surface of the substrate. Panels (a) and (b) share the same intensity scale. The analyte was optically interrogated from the top surface at an excitation power of 10 mW at 785 nm for an exposure time of 1 s with 20 accumulations.

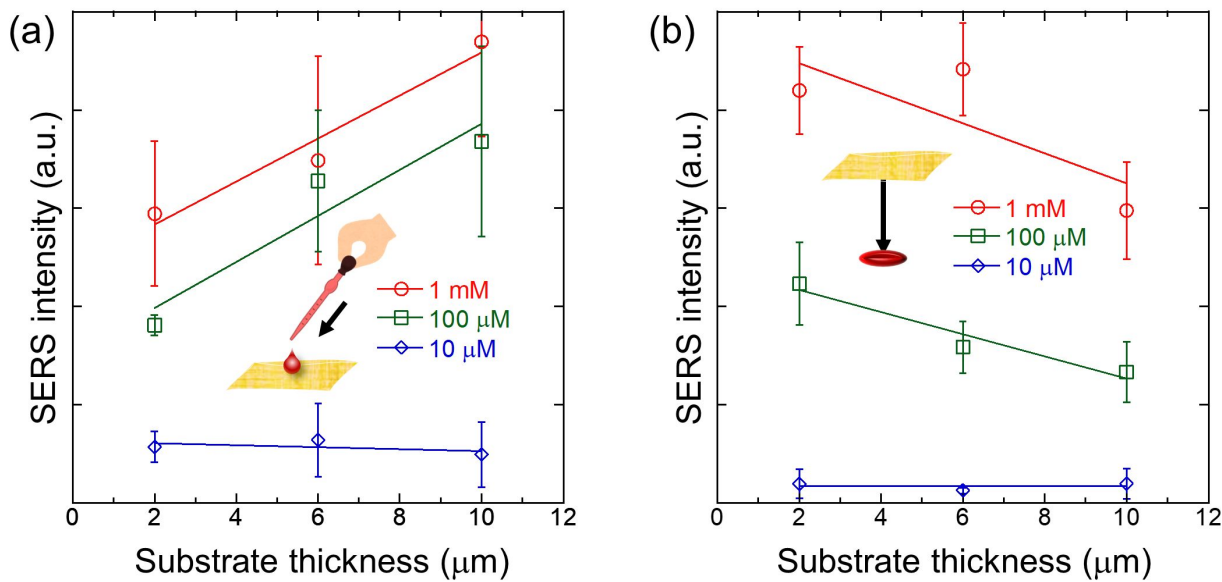

**Fig. 5** SERS peak intensity of an R6G aqueous solution at 775 cm$^{-1}$ obtained on the substrate with various substrate thickness values. Error bars indicate the standard deviations of repeated measurements (n = 3). (a) On the top surface of the substrate. Linear fitting was applied to visualize trends. (b) On the bottom surface of the substrate. Panels (a) and (b) share the same intensity scale. The analyte was optically interrogated from the top surface at an excitation power of 10 mW at 785 nm for an exposure time of 1 s with 20 accumulations.

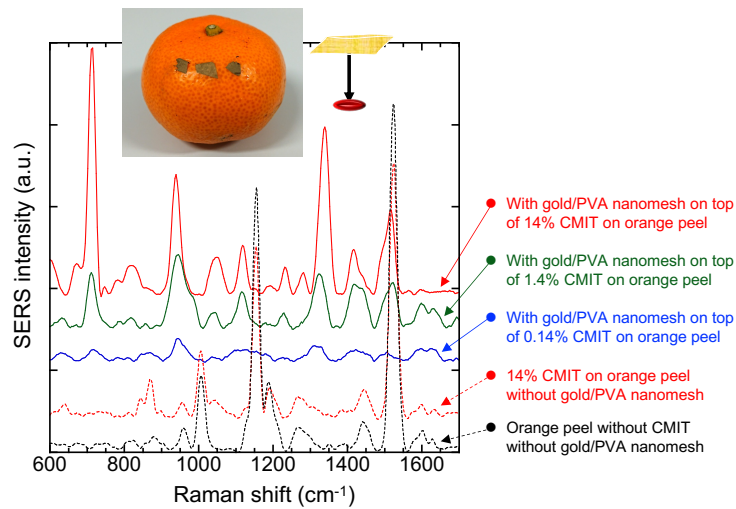

**Fig. 6** SERS spectra (solid cures) of an aqueous solution of CMIT at various concentrations on the bottom surface of the gold/PVA nanomesh substrate on an orange peel. The spontaneous Raman spectra of a 14% CMIT aqueous solution on an orange peel and the orange peel itself (dotted curves) are also shown. The analytes were optically interrogated from the top surface at an excitation power of 20 mW at 785 nm for a duration of 2.5 s.

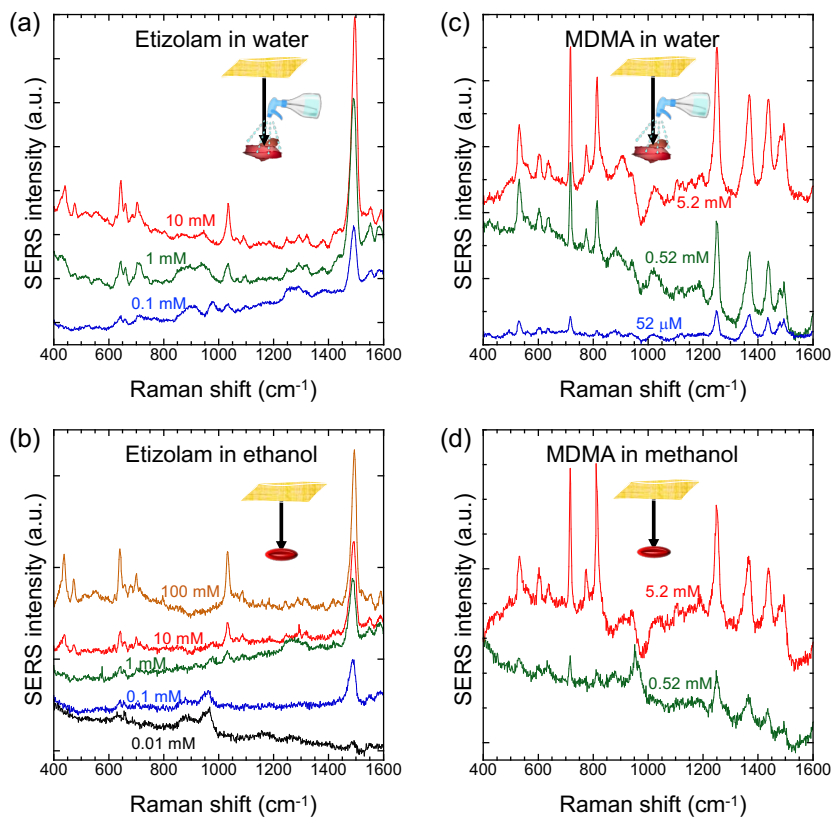

**Fig. 7** SERS spectra of etizolam and MDMA at various trace concentrations on an aluminium foil on the bottom surface of the gold/PVA nanomesh substrate. (a) Etizolam in water. (b) Etizolam in ethanol. (c) MDMA in water. (d) MDMA in methanol. The analytes were optically interrogated from the top surface at an excitation power of 10 mW at 785 nm for an exposure time of 1 s with 16 accumulations.